\documentclass[superscriptaddress, reprint, aps, showpacs, prb]{revtex4-2}

\usepackage[T1]{fontenc}
\usepackage{graphicx, xcolor, nicefrac}
\usepackage{soul}

\newcommand{\ie}{\emph{i.\,e.}}
\newcommand{\eg}{\emph{e.\,g.}}
\newcommand{\ea}{\emph{et al.}}

\newcommand{\co}{Co(II)}
\newcommand{\dc}{di-Co}
\newcommand{\tb}{\textit{tert}-butyl}

\newcommand{\ieap}{Institut f\"ur Experimentelle und Angewandte Physik, Christian-Albrechts-Universit\"at zu Kiel, 24098 Kiel, Germany}
\newcommand{\csic}{Centro de Física de Materiales CFM/MPC (CSIC-UPV/EHU), 20018 Donostia-San Sebastián, Spain}
\newcommand{\dipc}{Donostia International Physics Center (DIPC), 20018 Donostia-San Sebastian, Spain}
\newcommand{\ude}{Faculty of Physics and CENIDE, University of Duisburg-Essen, 47057 Duisburg, Germany}
\newcommand{\lis}{Centro de Química Estrutural, Institute of Molecular Sciences, Departamento de Química e Bioquímica, Faculdade de Ciências, Universidade de Lisboa, Campo Grande, 1749-016 Lisboa, Portugal}

\begin{document}

\author{Roberto Robles} \email{roberto.robles@csic.es} \affiliation{\csic}
\author{Chao Li} \affiliation{\ieap} 
\author{Sara Realista} \affiliation{\lis}
\author{Paulo Nuno Martinho} \affiliation{\lis}
\author{Manuel Gruber}  \affiliation{\ude}
\author{Alexander Weismann} \affiliation{\ieap} 
\author{Nicol\'as Lorente} \affiliation{\csic} \affiliation{\dipc}
\author{Richard Berndt} \affiliation{\ieap} %\email{berndt@physik.uni-kiel.de} 

\title{Interpreting tunneling spectroscopic maps of a dinuclear \co\ complex on gold}

\date{\today}

\begin{abstract}
Scanning tunneling microscope data from a dinuclear Co(II) complex adsorbed on Au(111) are analysed using density functional theory calculations.
We find that the interaction with the substrate substantially changes the geometry of the non-planar molecule.
Its electronic states, however, remain fairly similar to those calculated for a gas-phase molecule.
The calculations reproduce intriguing contrasts observed in experimental maps of the differential conductance $dI/dV$ and reveal the relative importance of geometric and electronic factors that impinge on the image contrasts.
For a meaningful comparison, it is important that the calculations closely mimic the experimental mode of measurement.
\end{abstract}

\maketitle

The scanning tunneling microscope (STM) ushered in the era of atomically resolved imaging~\cite{Binnig_1987}. 
However, the actual interpretation of STM images requires theoretical input.
Starting from some method to compute the electronic structure of the studied system, typical approaches can be divided into those that compute the actual electronic current~\cite{bardeen_tunnelling_1961,Sautet_1988,Sautet_1992,Cerda_1997,Hofer_2000,Hofer_2003,Blanco_2006, Duan2022}, and those that compute the local density of electron states (LDOS)~\cite{Tersoff1983,Tersoff1985,Chen_1990,Stokbro_1997,Lorente_2000,Olsson_2002,bocquet_theory_2009,lorente_stmpw_2019}.
This later approach is the Tersoff-Hamann approximation~\cite{Tersoff1983,Tersoff1985}, which assumes that the tip of a scanning tunneling microscope (STM) can be described by an $s$-wave and neglects tip-sample interactions, then the STM probes the LDOS at the position of the tip apex.
Although this interpretation is clear and simple, interpreting and understanding STM images of adsorbed molecules is, in some cases, not straightforward \cite{Blanco_2006, Zheng2021, Homberg_2023}.
Further complications arise when the $s$-wave approximation breaks down \cite{chen_tunneling_1990,Gross2011,mandi_chens_2015,Wang2018,
MartinezCastro2022,Duan2023}, \eg\ with functionalized tips \cite{Wagner2015, Gustafsson2017, Jelinek2017, Krejci2017, Abilio2024} or when inelastic  processes contribute a significant part to the tunneling current \cite{Schulz2015, Kuegel2018, Rolf2019, Reecht2020,  Li2024}.

One common tool to help the interpretation of STM experiments is to perform calculations of the geometric and electronic structure using density functional theory (DFT).
While this method is believed to accurately describe structural aspects, the energies of the obtained states, in particular unoccupied levels, tend to be less reliable.
In addition, calculating the widths of these states is a daunting task.
Finally, large molecules may require calculations for prohibitively large surface unit cells so that gas-phase calculations of the molecules are the only available substitute.
Depending on the mode of measurement used in experiments, \eg\ constant-current topographs, constant-height maps of the current, or maps of the differential conductance $dI/dV$ at either constant current or tip height, a number of states may contribute to the current to extents that are determined by the states' energies, widths, and spatial decays.
In view of the inherent uncertainties in the calculated results, a detailed comparison of experimental and calculated STM images is often avoided. The evolution of scanning-probe image in the context of predictive tools and artificial intelligence need to have reliable comparison between experiment and theory \cite{Kurki2024}.

Here, we analyze the intriguing case of the di-cobalt complex C$_{66}$H$_{86}$Co$_2$N$_4$O$_4$ (\dc, Figure~\ref{overview}a) adsorbed on Au(111) \cite{dicoone}.
Although the large size of the molecule does not exclude DFT calculations, the unit cell of a realistic coverage of molecules on the surface has to be simplified due to the extension of the herringbone-reconstruction of the Au(111) surface.
The structure of the complex is not planar in the gas phase and it retains some of its three-dimensionality on the surface.
This has a drastic influence on the image contrast.
In addition, the electronic states of the molecule are fairly sharp in the experimental data and may partially be disentangled by a comparison with calculated results.

Below, we first introduce the experimental and theoretical methods and highlight some characteristic experimental data.
In the main part of the article, the results of DFT calculations are presented.
We first discuss the structure of the adsorbed molecule and its electronic states.
Next, $dI/dV$ spectra are compared to calculated projected density of states (PDOS) and LDOS curves.
Finally, STM imaging modes (constant height, constant current, conductance mapping) are simulated and compared to the corresponding experimental data.

\section{Methods}

The \dc\ complex was synthesized according to the procedure of Ref.~\citenum{Shimakoshi2005}.
Au(111) single crystal surfaces were prepared by repeated Ar ion bombardment (1.5\,keV) and annealing to $450^\circ$\,C\@. 
The \dc\ molecules were sublimated from a heated crucible ($\approx 260^\circ$\,C) onto the substrate at $\approx25^\circ$\,C\@.
STM tips were electrochemically etched from W wire and annealed \textit{in vacuo}.
All STM experiments were carried out at $\approx 4.6$\,K and under ultrahigh vacuum conditions.
For measuring the differential-conductance $dI/dV$ a lock-in amplifier was used and a sinusoidal modulation voltage (0.5\,mV$_\mathrm{rms}$ at 667.8\,Hz) was applied to the sample voltage $V$\@.
The experimental $dI/dV$ spectra and STM images have been low-pass filtered.
In addition, the large-scale image has been corrected for a curved background.

To analyse the STM images,
electronic structure calculations were performed in the framework of DFT using the VASP code \cite{kresse_efficiency_1996}. Core electrons were treated using the projected augmented-wave (PAW) method \cite{kresse_ultrasoft_1999} and wave functions were expanded using a plane wave basis set with an energy cutoff of 400\,eV\@.
PBE was used as exchange and correlation functional \cite{perdew_generalized_1996}.
The missing van der Waals interactions in this functional were treated using the Tkatchenko-Scheffler method \cite{tkatchenko_accurate_2009}.
The description of the Co $d$-electrons was improved by using the LDA+U method as formulated by Dudarev \cite{dudarev_electron-energy-loss_1998} with the effective Hubbard $U$ value, $U-J=3$\,eV\@.

The Au(111) surface was simulated using the slab method with four atomic layers separated by a vacuum region of 21\,\AA\@. 
The coordinates of all atoms except the two bottom layers were relaxed until forces were smaller than 0.02\,eV/\AA\@.
The experimental unit cell (excluding the herringbone reconstruction) corresponds to a $12\times5$ rectangular unit cell and includes two \dc\ molecules. 
A matching simulation cell, as used in Ref.~\cite{dicoone}, includes 804 atoms and therefore it presents a formidable computational task. 
Here we used a smaller $7\times5$ rectangular unit cell including one \dc\ molecule.
In this cell the number of atoms is 442, which considerably reduces the computational cost of the simulations and allows for a more detailed analysis. 
We have compared the results using the $7\times5$ cell with simulations performed for the experimental unit cell, presented in Ref.~\cite{dicoone}. Despite the different positions and orientations of the molecules with respect to the substrate, we have found both results to be nearly identical.

STM simulations were done by applying the Tersoff-Hamann approximation \cite{Tersoff1985} using the method of Bocquet \ea\ \cite{bocquet_theory_2009} as implemented in the code STMpw \cite{lorente_stmpw_2019}.
A $(3\times3\times1)$ $k$-point sampling was used for the STM simulations, while due to the large size of the unit cell the geometry optimizations could be done using just the $\Gamma$-point. 
Charge transfers and magnetic moments were determined by Bader analyses \cite{tang_grid-based_2009}. 

An important analysis tool is the projected density of states (PDOS) onto atomic and molecular orbitals. The principle behind the PDOS is to analyze a complex electronic structure in terms of simpler wave functions, such as the ones corresponding to single-electron atomic or molecular orbitals.
Then the PDOS is just a density of states, where each electronic state of the system (given by the Bloch state $|n \vec{k}\rangle$ at energy $\epsilon_{n\vec{k}}$ because in the calculations we are using periodic boundary conditions) is also weighed by the overlap of the state with the atomic or molecular orbital that we are choosing to analyze our system.
Let us assume that $a$ is the atomic or molecular orbital, then the PDOS becomes:
\begin{equation}
    PDOS_a (E) = \sum_{n \vec{k}} |\langle a|n \vec{k}\rangle|^2 \delta (E-\epsilon_{n\vec{k}}).
    \label{eqPDOS}
\end{equation}

It is instructive to compare the above Eq.~(\ref{eqPDOS}) with the expression used to evaluate the $dI/dV$ in the calculations \cite{bocquet_theory_2009}, which is given by
\begin{equation}
    \frac{dI}{dV}(V)\propto \sum_{n \vec{k}} |\Psi_{n \vec{k}} (\vec{r}_0)|^2 \delta (eV-\epsilon_{n \vec{k}}),
\end{equation}
where $V$ is the applied bias, $\vec{r}_0$ is the tip's position over the surface, and $\Psi_{n \vec{k}} (\vec{r})$ is the wave function of the Bloch state $|n \vec{k}\rangle$ corresponding to the calculation's unit cell. Compared to the expression of the PDOS, Eq.~(\ref{eqPDOS}), the $dI/dV$ in the Tersoff-Hamann scheme corresponds to replacing the overlap of the Bloch state with the atomic or molecular orbital, by the spatial distribution of the Bloch wavefunction. 
This is a significant difference, because in the PDOS the spatial distribution of the state is integrated and does not depend on where the tip is placed.
The difference is even more drastic when we consider how the spatial distribution is on the vacuum-solid interface. Using the theory of Ref.~\cite{bocquet_theory_2009}, we can expand the wave function in its Fourier components along the surface (thanks to the unit-cell periodicity), and take the asymptotic behavior in the vacuum region along the coordinate $z$ ($z > 0$). Then, the wave function is given by:
\begin{equation}
        \Psi_{n,\vec{k}}(\vec{x},z)=\sum_{\vec{G}} C_{n \vec{k}}(\vec{G}) e^{-i (\vec{k}+\vec{G})\cdot \vec{x} } e^{-\xi(k,G,\epsilon_{n \vec{k}})|z|},
        \label{Fourier2}
\end{equation}
where $C_{n \vec{k}}(\vec{G})$ are the coefficients for the Fourier vector $\vec{G}$.
The function decays exponentially with the decay coefficient $\xi(k,G,E_{n,k})$, given by
\begin{equation}
        \xi(k,G,\epsilon_{n\vec{k}})=\sqrt{\frac{2 m^*(e\phi-\epsilon_{n\vec{k}})}{\hbar^2}+(\vec{k}+\vec{G})^2},
        \label{decay}
\end{equation}
where $\epsilon_{n\vec{k}}$ is the band energy for the state ${n\vec{k}}$, $m^*$ is the effective mass, and $e\phi$ is the work function. The decay in the vacuum region given by Eq.\,(\ref{decay}) is then a function of the work function ($e\phi$), but also of the \textit{spatial variations} along the surface given by the Fourier vectors, $\vec{G}$. As the tip recedes from the surface, $z$ increases, and large values of $(\vec{G})^2$ yield a very small value of the exponential in Eq. (\ref{Fourier2}), leading to a removal of fast spatially-varying features. This effect can greatly modify the final STM image that departs from simple addition of isosurfaces of molecular electronic structure or other simplifications like shown in Refs. \cite{Homberg_2023,olsson_stm_2003,Rolf,Grewal2024}.

\section{Experimental data}

\begin{figure}[h!]
	\includegraphics[width=1.0\columnwidth]{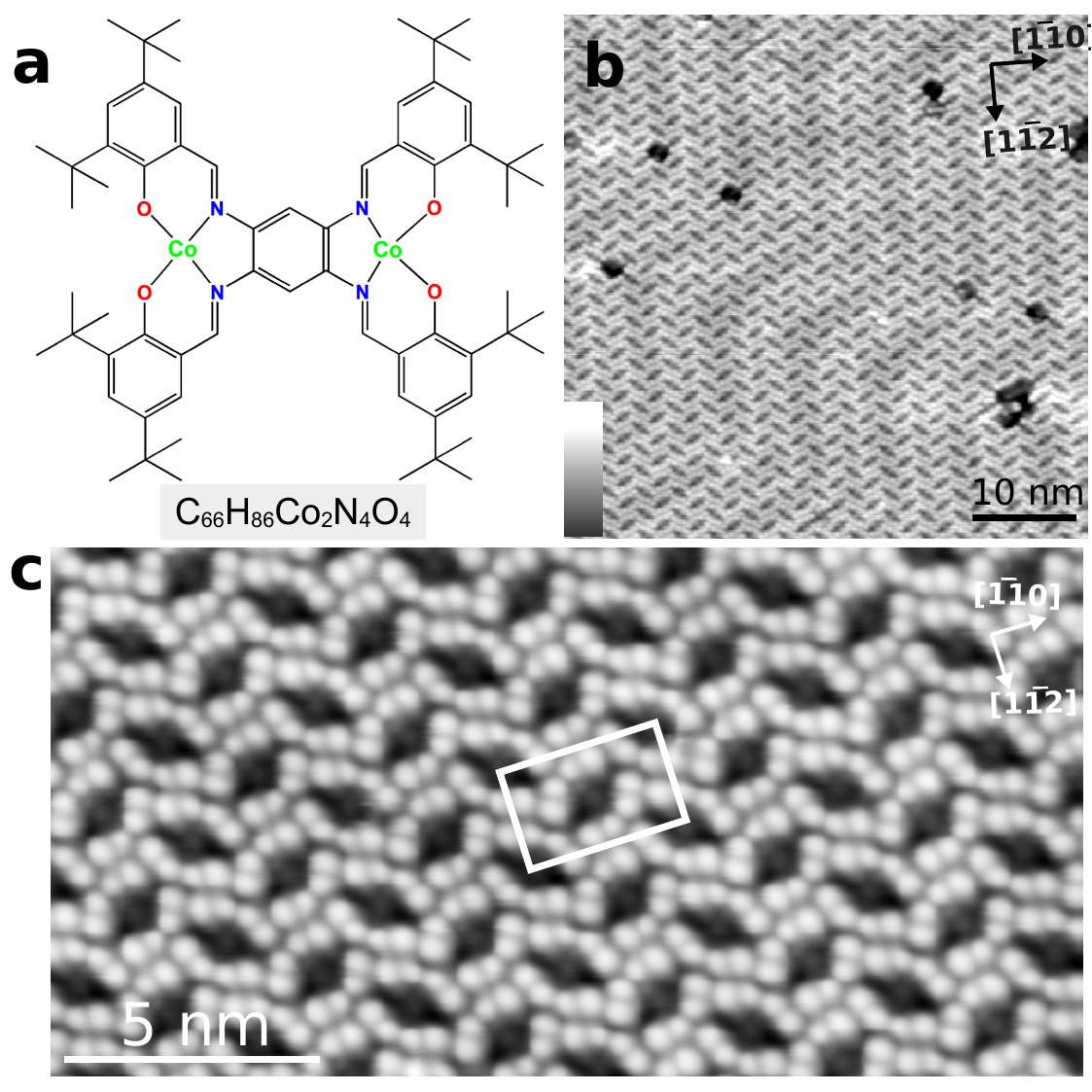}
	\caption{
	(a) Scheme of the investigated \dc~complex.
	(b) Topograph from a large monolayer island of \dc~on Au(111) recorded with $V=-1$\,V and $I=10$\,pA\@.
	The color scale used throughout the manuscript is displayed in the inset.
	The herringbone reconstruction of the Au substrate is discernible.	
	(c) Enlarged image of the molecular arrangement (0.1\,V, 140\,pA).
	The rectangular unit cell (edges 3.55 and 2.44\,nm) as determined from the STM images contains two molecules.
	It neglects the underlying reconstruction of the Au(111) substrate.
	The $[1\overline10]$ and $[11\overline2]$ directions of the substrate are marked with arrows. 
	\label{overview}}
\end{figure}

In the complex \dc, two Co(salophen) subunits are linked via a phenyl ring (Figure~\ref{overview}a).
The periphery of the complex is decorated with eight \tb\ moieties.
Deposition onto Au(111) surfaces at ambient temperature results in a rectangular pattern of the constant-current image(Figure~\ref{overview}b) with two molecules per unit cell (Figure~\ref{overview}c).
Eight major protrusions are observed per molecule, suggesting that the image features essentially indicate the bulky \tb\ subunits.

\begin{figure}[h!]
	\includegraphics[width=0.8\columnwidth]{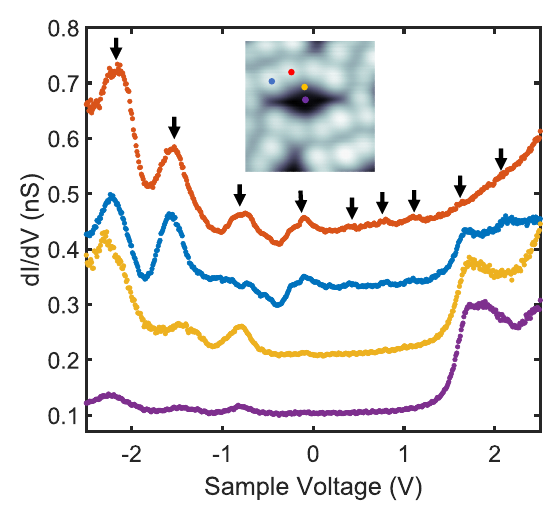}
	\caption{Spatially resolved spectra of the differential conductance $dI/dV$ at fixed heights.
	The feedback loop of the STM was disabled at $V= 2.5$\,V and $I=200$\,pA\@.
  The spectra were measured at the positions indicated in the topograph in the inset.
  They have been vertically shifted by 0.1, 0.2, 0.3, and 0.4~nS (bottom to top curves, respectively).
  Markers indicate voltages where reproducible features are observed.
	\label{spex}}
\end{figure}

Spectra of the differential conductance ($dI/dV$) covering a fairly wide voltage range reveal plenty of features, which may be expected for a molecule that couples weakly to the substrate (Figure~\ref{spex}).
In particular, several occupied states are resolved around the voltages $-2.2, -1.5, -0.8, -0.09$\,V and an unoccupied state is discernible at 1.7\,V with indication of weak features at 0.4, 0.8, 1.1, and 2.0\,V\@.
The data were measured at the tip positions indicated in the inset, which correspond to an outer \tb, an inner \tb, and a Co(II) center.
There are variations of the intensities and precise peak positions among the spectra, which may result from spatial variations of the relevant orbitals or of the vibrational excitation probabilities \cite{Pavlicek2013, Homberg2020}.

\begin{figure}[h!]
	\includegraphics[width=1.0\linewidth]{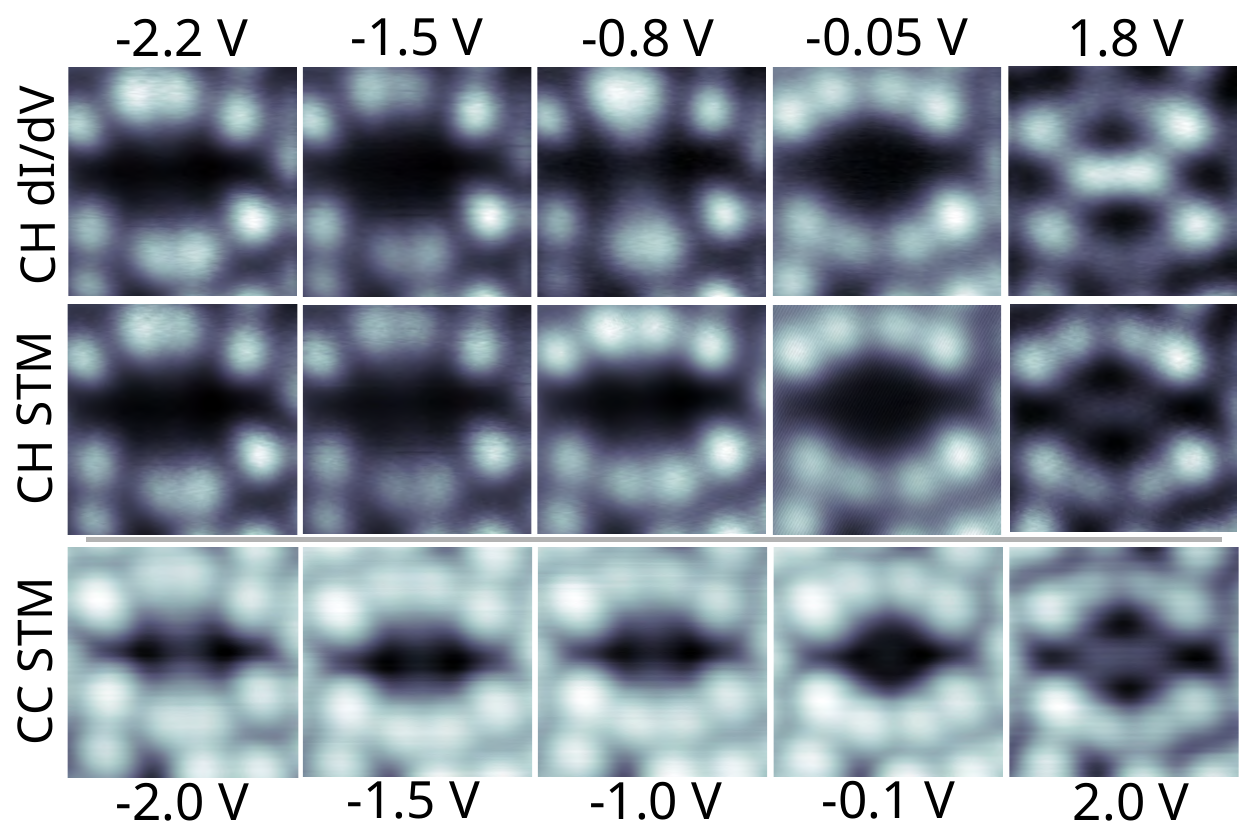}
	\caption{Bias dependent images of a single \dc\ complex within an ordered island.
	Top row: Maps of the differential conductance $dI/dV$ measured at constant heights determined opening the current feedback at $I=100$~pA and the voltages indicated at the top of the figure.
	Middle row: Constant height images recorded simultaneously with the constant $dI/dV$ maps.
	Bottom row: Constant current images recorded at $I=100$~pA and at the  voltages indicated at the bottom.
	\label{p1}}
\end{figure}

The main experimental results are displayed in Fig.~\ref{p1}, which combines series of constant-current images (CC STM), constant-height images (CH STM), and constant-height maps of $dI/dV$ (CH dI/dV) recorded from a molecule in an ordered island at various voltages.
In agreement with Fig.~\ref{overview}c, the lower row of Fig.~\ref{p1} shows the constant-current images with the previous eight bulky features at the position of the \tb\ moieties.
At the lowest and highest voltages used, some variations are observed. 
At $V=-2.0$\,V, the interior protrusions are hardly separable and at $V=2.0$\,V, these protrusions appear lower. 
Instead, a faint feature develops in the central area of the molecule. The middle row of Fig.~\ref{p1} shows the constant-height topography.
The protrusions in the constant-height images seem better resolved but do not add much further information.
As expected, the $dI/dV$ maps exhibit a more clear bias dependence (Fig.~\ref{p1}, upper row).
At $V=1.8$\,V, the central maximum is the most prominent feature of the image.
At negative voltages, a separation of the interior \tb\ features is resolved at $-0.05$ and at $-1.5$~ V but is absent at intermediate bias ($-0.8$\,V).

\section{Simulation results}

\subsection{Geometry and electronic states}

\begin{figure}[h!]
	\includegraphics[width=1.0\linewidth]{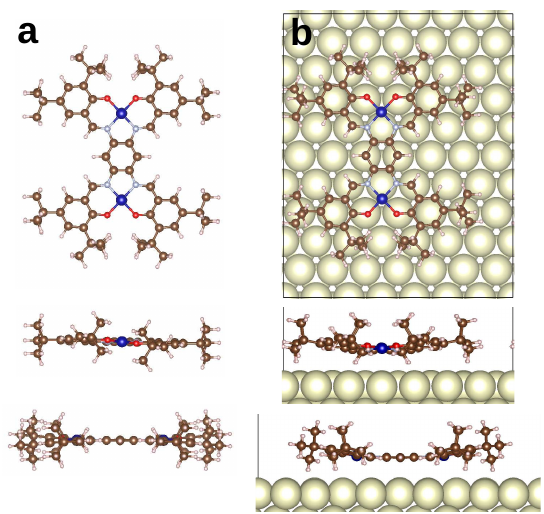}
	\caption{
 (a) Gas phase geometry of \dc\@.
The molecule is not fully symmetric due to steric interactions between the \tb\@ groups.
(b) Geometry of the molecule after deposition on Au(111). Lines show the unit cell used in the calculation.
\label{geometry}}
\end{figure}

We first performed calculations for the free-standing \dc\ molecule. Already in the gas phase the symmetry of the molecule is reduced due to steric interactions between the \tb\ groups, as can be seen in Fig.~\ref{geometry}a.
The molecule is magnetic, with each Co atom carrying a computed spin moment of 1.03\,$\mu_B$.
In the ground state, there is an antiferromagnetic (AF) coupling between the Co atoms, while the ferromagnetic (FM) solution is 2.4\,meV higher in energy. 
After deposition on the surface, the central backbone of the molecule bends to get closer to the surface due to the van der Waals interaction with the substrate (Fig.~\ref{geometry}b).
We estimate the bending by measuring the vertical distance (\ie\ the difference of the $z$ coordinates) between the central C atom and the lowest H atom.
This distance decreases from $\approx 2.5$\,\AA\ in the gas phase to $\approx 1.1$\,\AA\ on the surface.
Despite the considerable molecular distortion, the changes in the electronic structure are modest due to the limited hybridization with the surface that leads to a small charge transfer of 0.23 electrons from the molecule to the surface, and to maintaining the gas-phase magnetic properties. Upon adsorption, the spin moment of the Co atoms is 1.06\,$\mu_B$, with the AF configuration lying 3.0\,meV lower in energy than the FM one.

\subsection{Simulation of $dI/dV$ spectra}

Experimental $dI/dV$ spectra (Fig.\ref{spex}) are related to the local density of states (LDOS).
However, as first approximation $dI/dV$ spectra are sometimes compared to the projected density of states (PDOS). 
Fig.~\ref{PDOS}a shows the PDOS projected on the \dc\ states and on the orbitals of one of the Co atoms. 
Like in the experimental spectra, we find an abundance of peaks.
However, a direct comparison to the experimental $dI/dV$ spectra is difficult because the decay of the wavefunctions into vacuum is not taken into account.
It should also be noted that the energies of the correlated Co $d$ states, which are improved by the use of the LDA+U method \cite{dudarev_electron-energy-loss_1998}, depend on the choice of $U$\@.

\subsubsection{Local density of states}

\begin{figure}[h!]
	\includegraphics[width=0.95\linewidth]{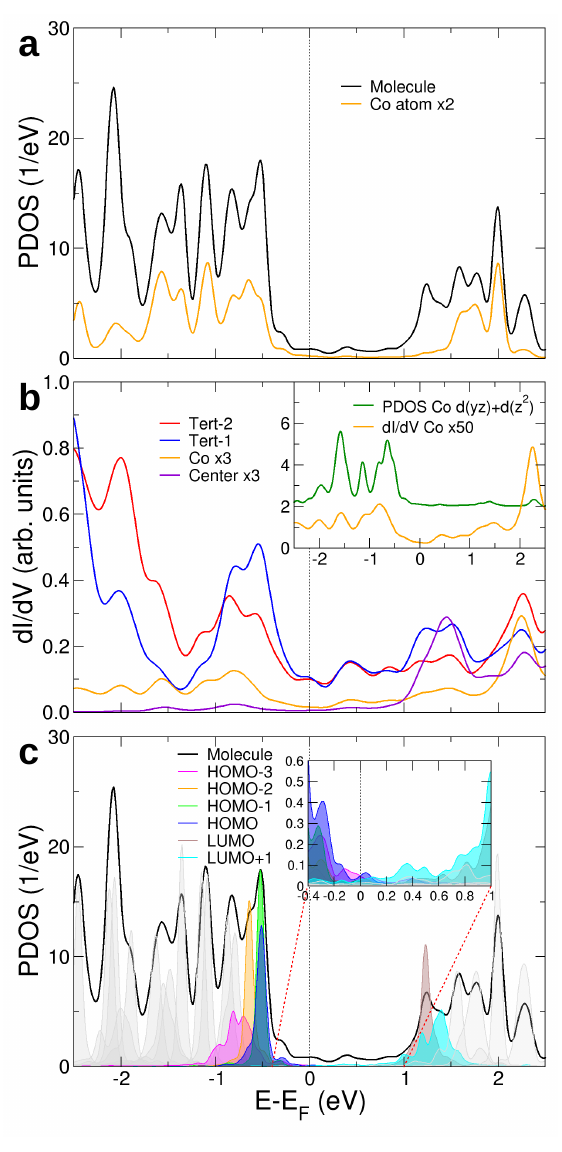}
	\caption{(a) DOS projected on \dc\ on Au(111) (black) and one of the Co atoms (orange).
(b) Calculated differential conductance $dI/dV$, using the LDOS, at a height of 21\,\AA\ above the metal surface.
The colors Tert-1 (blue), Tert-2 (red), Co (orange), and Center (violet) correspond to the experimental ones and the corresponding tip position in Fig.~\ref{spex}.
The inset shows $dI/dV$ over the Co atom along with the DOS projected on $d(yz)$ and $d(z^2)$ Co states.
(c) DOS projected on the MOs of the free-standing molecule in the adsorption geometry.
A broadening of 100\,meV has been applied to the results. MOs around the Fermi energy are highlighted with colors.
The inset shows a zoom around the gap region.
\label{PDOS}}
\end{figure}

The LDOS at 21\,\AA\ above the surface and calculated using the Tersoff-Hamann theory \cite{Tersoff1985} is displayed in Fig.~\ref{PDOS}b for selected horizontal positions. 
We verified that the same qualitative results are obtained for different heights.
Comparing with the projected DOS in Fig.~\ref{PDOS}a, some weak similarities can be found.
However, depending on the horizontal positions, different states that cause prominent PDOS features are observed in the LDOS high above the surface.
For example, the DOS over the center of the molecule (violet curve in Fig.\ref{PDOS}b) decreases faster than the DOS over the  \tb\ moieties (red and blue curves), similar to the experimental results.
Specially interesting is the slow decay of the DOS in the gap region ($\approx -0.4$\,eV to $\approx 1.0$\,eV) over the \tb\ moieties.
We can relate this behavior to the weak features of the experimental $dI/dV$ spectra.
In particular, the peak around the Fermi energy is also present in the experimental results.
As a further example, we compare the $dI/dV$ spectrum over Co (orange curve in Fig.~\ref{PDOS}b) with the projection on Co orbitals (orange curve in  Fig.~\ref{PDOS}a). 
Different peaks decay at different rates, resulting in different shapes of the $dI/dV$ and PDOS curves.
For instance, the strongest $dI/dV$ peak is located at $\approx$2.25~eV.
In contrast, the PDOS on Co only shows a minor feature at this energy.
Conversely, the strong PDOS peak at 2.00~eV is largely suppressed in the $dI/dV$ curve.
The reason can be traced back to the composition of both peaks.
The Co state at 2.25~eV is $d(yz)$, thus it has a $z$ component, and extends farther into the vacuum than the $d(xy)$ state at 2.00~eV\@.
The PDOS on the $d(yz)$ and $d(z^2)$ Co states (green curve in the inset of Fig.~\ref{PDOS}b) shares most peaks with the $dI/dV$ over Co showing that these orbitals are mainly responsible for the spectrum above Co.

\subsubsection{Projection on orbitals of the isolated molecule}

Fig.~\ref{PDOS}c shows projections of the states of the adsorbed complex onto some frontier molecular orbitals (MOs) (colored curves). The molecular orbitals correspond to the orbital $a$ appearing in the overlap on Eq. (\ref{eqPDOS}). In order to avoid geometrical spurious effects, the molecular orbitals are computed for the geometry of the adsorbed molecule and not of the free molecule.
The black curve is the DOS of the adsorbed molecule, \ie\ the sum of both spin directions shown in Fig.~\ref{PDOS}a.
The small width of the PDOS features reveals that most MOs exhibit little hybridization with the metal substrate.
However, there are exceptions like the HOMO--3 (magenta) and the LUMO+1 (light blue), which are significantly broader than the other projections.
We attribute the width of the HOMO--3 to the $d(yz)$ and $d(z^2)$ orbitals of the Co ion, which point not only into vacuum, but also toward the surface.
In the case of the LUMO+1, the hybridization is presumably caused by the large weight of the MO at the center of the molecule, which is close to the surface.

Most DOS peaks are not derived from a single orbital of the isolated molecule, rather they are combinations of several MOs.
For example, the highest occupied peak results mainly from the HOMO and the HOMO--1, and the lowest  unoccupied peak stems from the LUMO and the LUMO+1.

Returning to the $dI/dV$ spectra, we find that the projection onto the LUMO+1 (light blue) closely resembles the $dI/dV$ spectrum above the center of the molecule (Fig.~\ref{PDOS} b, violet).
The inset of Fig.~\ref{PDOS}c shows that the tails of several MOs extend into the gap region.
This effect does not only involve the HOMO and the LUMO, but is also due to the strongly hybridized states HOMO--3 and LUMO+1. 

\begin{figure}[h!]
	\includegraphics[width=1.0\linewidth]{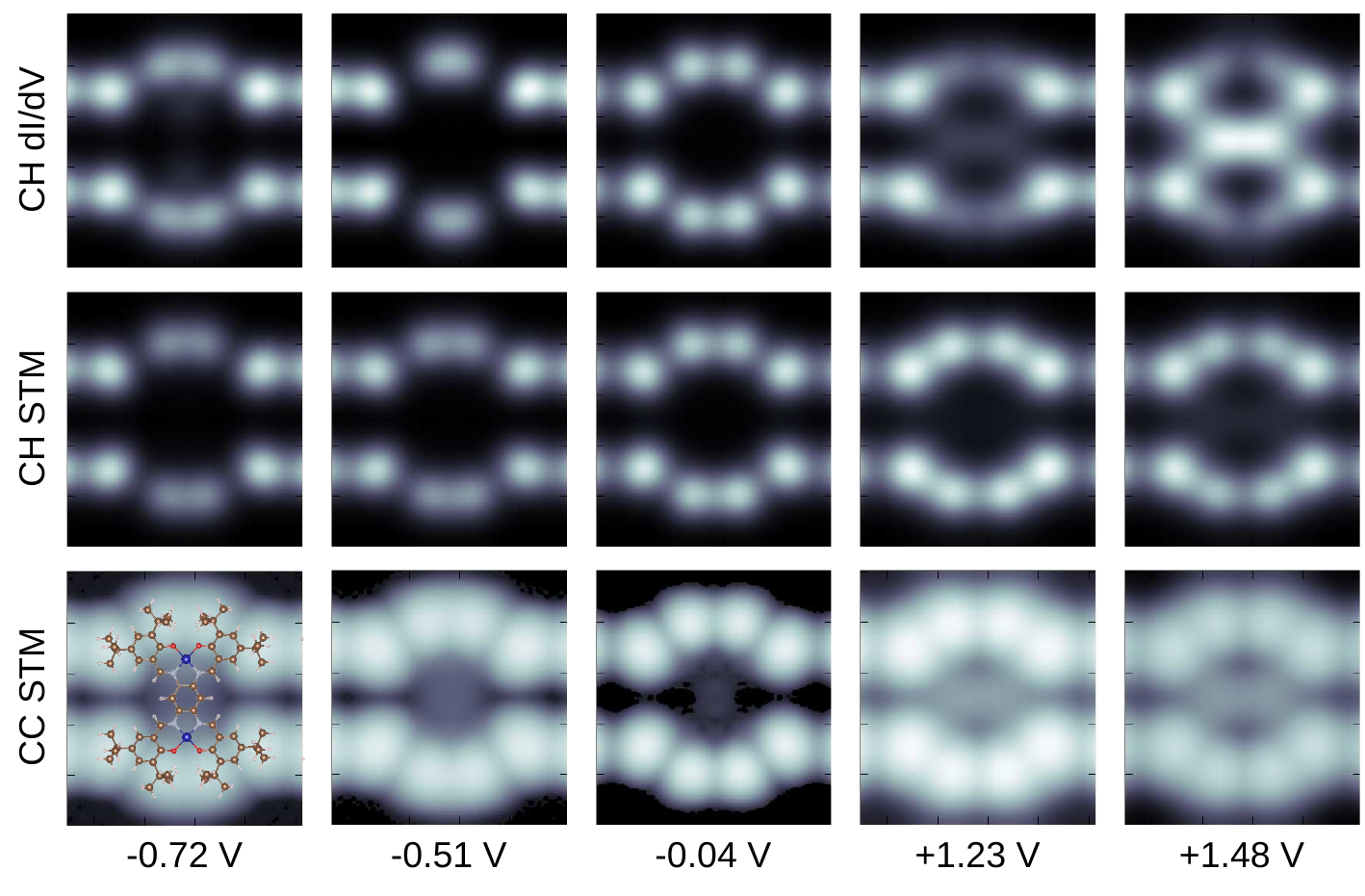}
	\caption{Simulated images of a \dc\ molecule at the indicated voltages.
Top row: Maps of the differential conductance $dI/dV$ at constant height.
Middle row: Current images at constant height.
Bottom row: Constant current images.
The optimized geometry of the molecule is overlaid at the bottom left corner.
	\label{maps}}
\end{figure}

\subsection{Simulation of images}

Using STMpw \cite{lorente_stmpw_2019}, maps of $dI/dV$ at constant height, current images at constant height, and constant current images were calculated as a function of the sample voltage.
Selected results are shown in Fig.~\ref{maps}.
Like in the experimental data, the simulated mode of measurement drastically affects the appearance of the images.
$dI/dV$ maps show the most pronounced bias dependence as expected.
Moreover, the constant current mode enhances small conductances and therefore seems to broaden the molecular features compared to the current image.

The simulated images agree remarkably well with the experimental data of Fig.~\ref{p1}.
All imaging modes lead to eight prominent features at the positions of the \tb\ moieties.
In the $dI/dV$ maps, their shapes vary with the voltage.
In particular, the inner protrusions are almost fused at $V=-0.72$ and $-0.51$\,V, are clearly resolved at $-0.04$\,V and are only weakly discernible at the shown positive voltages.
Moreover, a strong feature is resolved in the central area at 1.48\,V, while its intensity is very low at $-0.72$ and 1.23\,V\@.
These evolutions can also be recognized in the experimental $dI/dV$ maps, albeit at slightly different voltages.
The experimental maps at $1.8, -0.05, $ and  $-0.8$\,V resemble the simulated ones at 
$1.48, -0.04$, and $-0.72$\,V\@. 

\begin{figure}[h!]
	\includegraphics[width=1.0\linewidth]{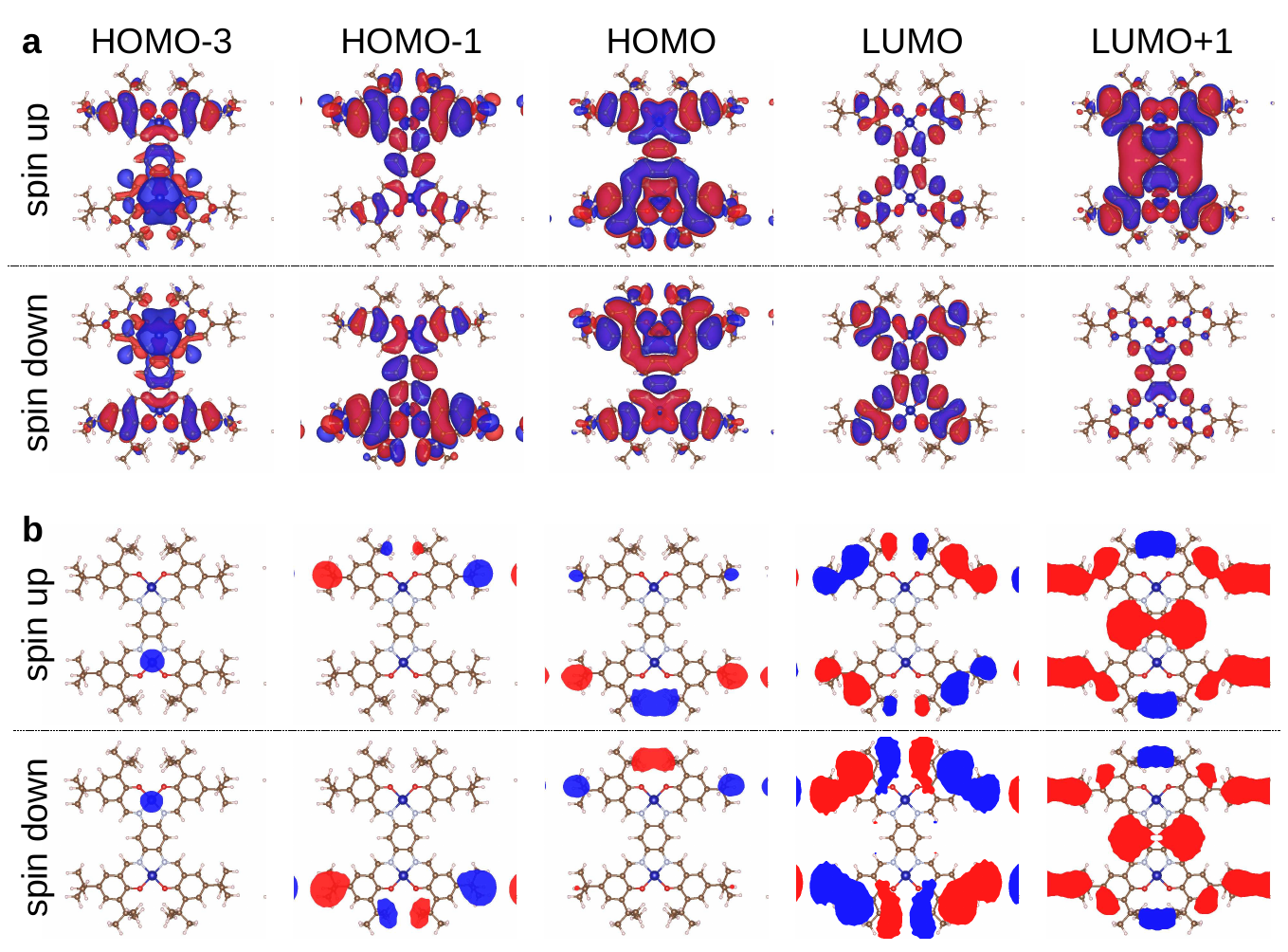}
	\caption{(a) Isosurfaces of the wavefunctions of selected MOs of the gas-phase \dc\ molecule with the geometry of the adsorbed case, the two colors refer to two phases ($\pm$) in the wavefunction.
	Spin up and down MOs are shown as marked by the label on the left.
	(b) Planar sections at $\approx$ 4.5~\AA\ above the molecular plane.
	\label{mos}}
\end{figure}

\subsubsection{Energy dependence of the imaged states}
Next, we analyze the origins of some features of the STM images.
Starting with the $-0.51$\,V simulation, in which the interior protrusions are fused, 
we find two main contributions to the image contrast from the projected DOS in Fig.~\ref{PDOS}c, namely the HOMO and the HOMO--1.
The corresponding wavefunctions of the gas-phase molecule are shown in Fig.~\ref{mos}a along with the geometry of the adsorbed molecule.
The HOMO and the HOMO--1 are hardly spin-polarized and consequently the two spin directions have similar spatial distributions and contributions to the image contrast.
However, a direct comparison of the isosurfaces with the experimental data, in particular in the constant-height $dI/dV$ maps that present the clearest information on the electronic structure, is not illuminating.
The contrasts in the images of \dc\ are predominantly caused by the three-dimensional geometry of the molecule, and therefore the variations in the isosurfaces of the different MOs are hardly seen in the STM maps. 
Rather than eight protrusions, the isodensity surfaces show plenty of additional details.

\subsubsection{Spatial dependence of the imaged states}

To more closely match the measurement process, we use planar sections of the wavefunctions at $\approx$ 4.5~\AA\ from the molecular plane (Fig.~\ref{mos}b).
In this way, their evolutions in vacuum, which depend on the electronic structure and on the geometry as well, are taken into account.
Considering both spin up and spin down, the sections of the HOMO and the HOMO--1 feature eight protrusions as observed in the experiment.
The fusion of the interior protrusions is caused by the HOMO\@.
Its wavefunction does not change phase between the interior \tb\ groups, a nodal plane is absent, and for this reason the interior groups appear fused at large heights.

The experimental $dI/dV$ map at $-0.8$\,V displays fused interior protrusions, but in addition a maximum is present over the Co ions.
A corresponding signal may be found in the simulated $dI/dV$ map at $-0.72$\,V\@.
Figure~\ref{PDOS}c reveals that several MOs are relevant in this case.
Some of them, like HOMO--2 or HOMO--4 (not shown), do not add additional features beyond the eight \tb\ protrusions. 
The HOMO--3, however, does exhibit a feature over the Co atom.
It is due to the $d(yz)$ and $d(z^2)$ states of the Co ion, which extend far into the vacuum. 

Another prominent feature of the STM images is the maximum above the central region of the molecule.
In the simulations, it is particularly strong at 1.48\,V, but is also observed at 1.23\,V\@.
According to Figure~\ref{PDOS}c, the most important orbitals at the latter energy are the LUMO and the LUMO+1.
The LUMO+1 has an important weight on the central part and therefore gives rise to the signal at 1.23\,V\@.
At 1.48\,V, a further contribution from the LUMO+2 (not shown) adds to the signal.
This spatial pattern of this orbital and the LUMO+1's resemble each other, resulting in a stronger feature at the molecular center.

Finally, we address the images at $-0.04$\,V\@. This energy falls inside the HOMO-LUMO gap, and in many cases there are 
not enough molecular states to allow a good simulation of the STM image. Here the tails of several MOs extend into this 
energy (see inset of Fig.~\ref{PDOS}c), and their combination leads to images with the eight protrusions above the 
molecule as observed in the experimental low-bias maps.

\section{Conclusions}

The  non-planar \dc\ complex adsorbed on Au(111) was investigated with STM and DFT calculations.
Despite the complexity of the molecule and its three-dimensional geometry the experimental and the theoretical results using the Tersoff-Hamann approach match fairly well.
While projected densities of states and isodensity surfaces, which are readily available in various DFT codes, are useful for discussing the electronic structure of the molecule on the surface in terms of its gas-phase orbitals, they fail to match experimental data.
To analyse constant-current images, current images, or maps of the differential conductance, it is necessary to take the spatial decay of the relevant wave functions into account and to closely mimic the mode of measurement used in the experiments.
In the present case of \dc, much of the image contrast is caused by the geometric height of the molecular subunits together with the extension of the molecular orbital into the vaccum region \cite{olsson_stm_2003}. Additionally, the filtering effect of the different decay of Fourier components of the electronic structure leads to non-trivial final images \cite{Homberg_2023, Rolf}.
However, Co $d(yz)$ and $d(z^2)$ states may also be detected in energy resolved data because they lead to sharp spectral features.

\subsection*{Acknowledgement}
RR and NL thank projects  PID2021-127917NB-I00 by MCIN/AEI/10.13039/501100011033, QUAN-000021-01 by the Gipuzkoa Provincial Council, IT-1527-22 by the Basque Government, 202260I187 by CSIC, ESiM project 101046364 by the EU, and computational resources by Finisterrae III (CESGA). Views and opinions expressed are however those of the author(s) only and do not necessarily reflect those of the EU. Neither the EU nor the granting authority can be held responsible for them.
C.L. thanks the Alexander von Humboldt Foundation for a Research Fellowship for Postdoctoral Researchers and also acknowledges support from Kiel Nano, Surface and Interface Science (KiNSIS). 
Centro de Química Estrutural (CQE) and Institute of Molecular Sciences (IMS) acknowledge the financial support of Fundação para a Ciência e Tecnologia (FCT), projects UIDB/00100/2020 (https://doi.org/10.54499/UIDB/00100/2020), UIDP/00100/2020 (https://doi.org/10.54499/UIDP/ 00100/2020), and LA/P/0056/2020 (https://doi.org/10.54499/LA/P/0056/2020), respectively.  P.N.M. and S.R. thank FCT for the research contracts CEEC-IND/00509/2017 (https://
doi.org/10.54499/CEECIND/00509/2017/CP1387/ CT0029) and 2020.02134.CEECIND (https://doi.org/10.54499/2020.02134.665/CEECIND/ CP1605/CT0002). M.G. acknowledges funding from the Deutsche Forschungsgemeinschaft (DFG, German Research Foundation) Project No. 278162697-SFB 1242.

%\subsection*{Author contributions}
%SR and PNM synthesized the molecules.
%CL performed the STM experiments.
%RR and NL did the DFT calculations.
%RB and RR analysed the results and wrote the manuscript.
%All authors commented on the manuscript.

%\subsection*{Data availability}
%Raw data may be obtained from the corresponding authors upon reasonable request.

\bibliography{dico}

\end{document}